# *New approach method for solving nonlinear differential equations of blood flow with nanoparticle in presence of magnetic field*

M.Hamzeh
Department of Mechanical Engineering
K. N. Toosi University of Technology
Tehran, Iran
m.hamzeh@mail.kntu.ac.ir

A.Kachabi, M.Heydari sipey, D.D.Ganji
Department of Mechanical Engineering
Noshirvani University of Technology
Babol, Iran
mirgang@nit.ac.ir

*Abstract*

In this paper, effect of physical parameters in presence of magnetic field on heat transfer and flow of third grade nonNewtonian Nanofluid in a porous medium with annular cross sectional analytically has been investigated. The viscosity of Nanofluid categorized in 3 model include constant model and variable models with temperature that in variable category Reynolds Model and Vogel's Model has been used to determine the effect of viscosity in flow filed. analytically solution for velocity, temperature, and nanoparticle concentration are developed by Akbari-Ganji's Method (AGM) that has high proximity with numerical solution (Runge-Kutta 4th-order). Physical parameters that used for extract result for non-dimensional variables of nonlinear equations are pressure gradient, Brownian motion parameter, thermophoresis parameter, magnetic field intensity and Grashof number. The results show that the increase in the pressure gradient and Thermophoresis parameter and decrease in the Brownian motion parameter cause the rise in the velocity profile. Also the increase in the Grashof number and decrease in MHD parameter cause the rise in the velocity profile. Furthermore, either increase in Thermophoresis or decrease in Brownian motion parameters results in enhancement in nanoparticle concentration. The highest value of velocity is observed when the Vogel's Model is used for viscosity.

I. INTRODUCTION

The applications of NonNewtonian fluids are broad in many industrial and technology, such as drug delivery, biochemical reactions, melts of polymers, biological solutions, paints, tars, asphalts, and glues. Recent advances in nanotechnology have led to an increasing interest in research about Nanofluids but so far nonNewtonian Nanofluids have received very little attention [1]. One of the most important multicomponent compounds is blood that consists of red and white blood cells, plasma, platelets, and the etc. In review of NonNewtonian Nanofluids importance an experimental study Praveen Kumar et al. [2] investigated the effect of gold nanoparticles in blood from biomedically view point which can be used in drug delivery applications. the flow of Newtonian and nonNewtonian fluids through two infinite parallel vertical plates has been investigated by numerous authors. Ellahi et al. [3] used Reynolds and Vogel's model for viscosity of nonNewtonian Nanofluid in a porous medium, also in other work Ellahi et al. [4] studied the effect of MHD[1] on nonNewtonian Nanofluid flow between two coaxial cylinders. NonNewtonian fluids have many applications in biomedical science. The physical properties of blood as a nonNewtonian fluid is presented by Abdel Baieth [5]. Ogulu and Amos [6] modeled pulsatile blood flow in the cardiovascular system employing the Navier-Stokes equation. Investigation on nonNewtonian fluid flow in microchannels and flow characteristics of deionized water and the PAM solution over a wide range of Reynolds numbers has been done by Tang et al. [7]. Yoshino et al. [8] presented a new numerical method for incompressible nonNewtonian fluid flows based on the LBM[2]. Their simulations indicate that the method can be useful for practical nonNewtonian fluid flows, such as shear-thickening (dilatant) and shear-thinning (pseudo-plastic) fluid flows. Xu and Liao [9] studied the unsteady MHD viscous flows of nonNewtonian fluids caused by an impulsively stretching plate by means of homotopy analysis method to investigate the effect of integral power-law index of these nonNewtonian fluids on the velocity. In an experimental study Kumar et al. [10] investigated the effect of gold nanoparticles in blood from biomedically viewpoint. Many studies are focused on second and third grade nonNewtonian fluids which some of them are presented in this section. Modeling and solution of the unsteady flow of an incompressible third grade fluid over a porous plate within aporous medium is investigated by Aziz et al. [11]. The MHD flow due to non-coaxial rotations of a porous disk, moving with uniform acceleration inits own plane and a second grade fluid at infinity is examined by Asghar et al. [12]. Keimanesha et al. [13] solved the problem of a third grade nonNewtonian fluid flow between two parallel plates by Ms-DTM[3]. Hayat et al. [14-16] completely discussed the treatment of third grade non-Newtonian fluids in different applications. Hatami and Ganji [17] analytically investigated Heat transfer and flow analysis for SA-TiO2 nonNewtonian Nano-fluid passing through the porous media between two coaxial cylinders. M.Hatami et al. [18] Investigated third-grade non-Newtonian blood flow in arteries under periodic body acceleration using multi-step differential transformation method.

---

[1] magnetohydrodynamic
[2] lattice Boltzmann method

To solve differential equations, there are simple, accurate methods known as weighted residual methods (WRMs). Collocation, Galerkin and Least Square are examples of the WRMs. In this context Hatami et al. [19] studied the third grade non-Newtonian blood conveying gold nanoparticles in a porous and hollow vessel by two analytical methods called Least Square Method (LSM) and Galerkin Method (GM). Also Hatami et al. [20] used Collocation Method and (LSM) for Heat transfer and flow analysis for SA-TiO2 nonNewtonian Nano-fluid passing through the porous media between two coaxial cylinders. Akbari N et al. [21] by Flex PDE Software and (CM) investigated Blood Flow with Nanoparticles in a Magnetic Field as a Third Grade NonNewtonian Through Porous Vessels.

Recently several researchers have been done attempts to solve equations with Akbari-Ganji's Method (AGM) [26]. In this paper, the AGM is used to solving the nonlinear heat transfer and fluid flow equations. Some advantages of this method such as being a simple, concise and innovative method, demonstrate the reliability of AGM compared with other methods. In the following the comparison between AGM, WRMS and numerical method, will show the accuracy of AGM and will endorse the purport that have been proposed. Viscosity of Nanofluid is determined using Constant Model, Reynolds' Model and Vogel's Model. Also the effects of some parameters such as Brownian motion and thermophoresis parameters on velocity and temperature and Nanofluid concentration profiles are investigated.

## II. GOVERNING EQUATIONS

Consider a steady, incompressible, nonNewtonian Nanofluid in presence of magnetic field and in the porous media of hollow vessel as shown in Fig. 1. For modeling such a structure, the Nanofluid's density, $\rho$ should be defined as:

$$\rho = \phi \rho_p + (1-\phi)\rho_{fo} \cong \phi \rho_p + (1-\phi)[\rho_f (1-\beta_T(\theta - \theta_\omega))] \quad (1)$$

Where $\rho_{fo}$ is the base fluid's density, $\theta_\omega$ is a reference temperature, $\rho_f$ is the base fluid's density at the reference temperature, $\beta_T$ is the volumetric coefficient of expansion. Taking the density of base fluid as that of the Nanofluid, as adopted by Yadav et al. [29], the density $\rho$ in (1), thus becomes:

$$\rho \cong \phi \rho_p + (1-\phi)[\rho_f (1-\beta_T(\theta - \theta_\omega))] \quad (2)$$

$\rho_f$ is the Nanofluid's density at the reference temperature [23]. Clearly a viscous fluid is governed by continuity and Navier–Stokes equations and when the fluid is considered to be incompressible, the conservation of momentum, total mass, thermal energy, and nanoparticles, are as follows [3],

$$\rho_f (\frac{\partial V}{\partial t} + V.\nabla) = div T - \frac{\mu \varphi}{\kappa}(1 + \mu \lambda_r \frac{\partial}{\partial t})V \quad (3)$$
$$+\{\phi \rho_p + (1-\phi)[\rho_f (1-\beta_T(\theta - \theta_\omega))]\}g + J \times B$$

$$(\rho c)_f (\frac{\partial \theta}{\partial t} + V \cdot \nabla \theta) = \kappa \nabla^2 \theta \quad (4)$$
$$+(\rho c)_P [D_b \nabla \phi \cdot \nabla \theta + \frac{D_T}{\theta_\omega} \nabla \theta \cdot \nabla \theta]$$

$$(\frac{\partial \phi}{\partial t} + V \cdot \nabla \phi) = D_b \nabla^2 \phi + \frac{D_T}{\theta_\omega} \nabla^2 \theta \quad (5)$$

where $\phi$ is the nanoparticles volume fraction and $\varphi$ is porosity of medium. Since we considered electrically conducting fluids, so the fourth term on the right hand side of (3) is considered. The following equation gives the origin of B (Ohm's law) [12]:

$$J = \sigma(E + V \times B) \quad (6)$$

in which $E$ is the electric field. In the last term of (6) on the right hand side $B = B_0 + b$ is the total magnetic field, is the electrical conductivity, $V$ is the velocity vector and $J$ is the electric current density. For small magnetic Reynolds number the induced magnetic field is neglected and hence we can easily write [11,12]

$$J \times B = -\sigma B_0^2 V \tag{7}$$

---
[3] Multistep Differential Transformation Method

Although this problem can be solved for each boundary values, but in this study, it's assumed that each variable, $v$, $\theta$, and $\phi$, has an initial value in the first boundary $r = R_1$, and its value reaches to zero in the second boundary $r = R_2$. Along with this boundary conditions [3],

$$v(R_1) = v_0 \qquad \theta(R_1) = \theta_\omega \qquad \phi(R_1) = \phi_\omega$$

$$v(R_2) = 0 \qquad \theta(R_2) = 0 \qquad \phi(R_2) = 0 \tag{8}$$

Stress in a third-grade fluid is given by

$$T = -pI + \mu A_1 + \alpha_1 A_2 + \alpha_2 A_1^2 + \beta_1 A_3 + \beta_2(A_1 A_2 + A_2 A_1) + \beta_3(trA_1^2)A_1 \tag{9}$$

where $\alpha_1, \alpha_2, \beta_1, \beta_2$ and $\beta_3$ are the material modules and are considered to be functions of temperature in general. In (9), $-pI$ shows the spherical stress due to the restraint of incompressibility, and the kinematical tensors $A_1$, $A_2$ and $A_3$ can be defined by following equations [4]

$$A_1 = (\nabla V) + (\nabla V)^t \tag{10}$$

$$A_n = \frac{dA_{n-1}}{dt} + A_{n-1}(\nabla V) + (\nabla V)^t A_{n-1}, n = 2,3 \tag{11}$$

Where $V = [0, 0, v(r)]$ denotes the velocity field, grad is the gradient operator, and $D/Dt$ is the material time derivative, which is defined by

$$\frac{D(\cdot)}{Dt} = \frac{\partial(\cdot)}{\partial t} + [\nabla(\cdot)]V \tag{12}$$

When the fluid is locally at rest, the specific Helmholtz free energy is minimum [24].

$$\mu \geq 0 \qquad \alpha_1 \geq 0 \qquad |\alpha_1 + \alpha_2| \leq \sqrt{24\mu\beta_3}$$
$$\beta_1 = \beta_2 = 0, \beta_3 \geq 0$$
$$\tag{13}$$

Ellahi et al. [3] considered that the fluid is thermodynamically compatible, and therefore (9) reduced to

$$T = -pI + \mu A_1 + \alpha_1 A_2 + \alpha_2 A_1^2 + \beta_3(trA_1^2)A_1 \tag{14}$$

(4) can be rearranged as

$$T = -pI + [\mu + \beta_3(trA_1^2)]A_1 + \alpha_1 A_2 + \alpha_2 A_1^2 \tag{15}$$

By introducing the following non-dimensional parameters

$$\bar{v} = \frac{v}{V_0} \qquad \bar{r} = \frac{r}{R} \qquad \bar{\mu} = \frac{\mu}{\mu_0}$$
$$\bar{\theta} = \frac{\theta - \theta_\omega}{\theta_m - \theta_\omega} \qquad \bar{\phi} = \frac{\phi - \phi_\omega}{\phi_m - \phi_\omega} \tag{16}$$

Where, $V_0$, $\mu_0$, $\theta_\omega$, $\theta_m$, and $\phi_m$ denote the reference velocity, viscosity, pipe temperature, fluid temperature, and mass concentration, respectively. Substituting (15) in the balance of linear momentum and using the non-dimensional quantities given in (16), the dimensionless forms of the governing (3) – (5), after dropping bars for simplicity, lead to the following nondimensional coupled equations [4],

$$\frac{d\mu}{dr}\frac{dv}{dr} + \frac{\mu}{r}\frac{dv}{dr} + \mu\frac{d^2v}{dr^2} + \frac{A}{r}(\frac{dv}{dr})^3 + 3A(\frac{dv}{dr})^2\frac{d^2v}{dr^2}$$
$$= Pv + M^2 v + c - G_r \theta - B_r \phi \tag{17}$$

$$\alpha\frac{d^2\theta}{dr^2} + \frac{1}{r}\frac{d\theta}{dr} + N_b\frac{d\theta}{dr}\frac{d\phi}{dr} + \alpha_1 N_t(\frac{d\theta}{dr})^2 = 0 \tag{18}$$

$$N_b(\frac{d^2\theta}{dr^2} + \frac{1}{r}\frac{d\theta}{dr}) + N_t(\frac{d^2\phi}{dr^2} + \frac{1}{r}\frac{d\phi}{dr}) = 0 \tag{19}$$

where $P$ is porosity and $M$ is MHD parameters. With the boundary conditions as the form,

$$v(1) = 1 \qquad \theta(1) = 1 \qquad \phi(1) = 1$$
$$v(2) = 0 \qquad \theta(2) = 0 \qquad \phi(2) = 0 \tag{20}$$

Equations (17) – (19) is defined by considering the following parameters [4],

$$A = \frac{2\beta_3 v_0^2}{\mu_0 R^2}, c = \frac{(\partial P/\partial Z)R^2}{\mu_0 v_0}, P = \frac{\varphi}{\kappa R^2}, M^2 = \frac{\sigma B_0^2 R^2}{\mu_0}$$
$$G_r = \frac{(\theta_m - \theta_\omega)\rho_{f,\omega}R^2(1-\phi_\omega)g}{\mu_0 v_0}, N_t = \frac{D_T(\theta_m - \theta_\omega)}{\theta_\omega}$$
$$N_b = D_b(\phi_m - \phi_\omega), B_r = \frac{(\rho P - \rho\omega)R^2(\phi_m - \phi_\omega)g}{\mu_0 v_0}$$
$$\tag{21}$$

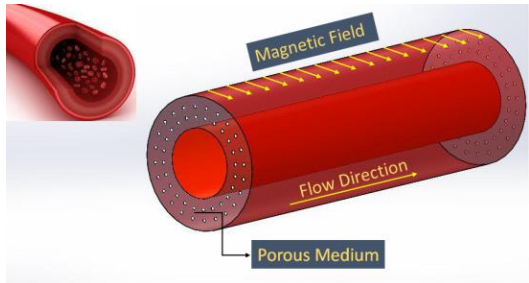

Figure 1. The geometry of problem

### III. DESCRIBE THE AKBARI-GANJI'S METHOD (AGM) AND APPLIED TO PROBLEM

The general form of the steady one dimensional heat transfer differential equation is defined as follows:

$$f(\frac{d^2\theta}{dr^2}, \frac{d\theta}{dr}, \theta, q_r) = 0 \qquad (22)$$

And the boundary conditions are as follows:

$$r = r_0 \Rightarrow \theta = \theta_0$$
$$r = r_L \Rightarrow \theta = \theta_L \qquad (23)$$

In The AGM, a total answer with constant coefficients is required in order to solve the differential equation. There's a point in chosen answer, and that is number of coefficients of answer must be multiple of order of differential equation (number of boundary condition).

Now, we choose the answer for differential equation as follows:

$$\theta(r) = \sum_{i=0}^{n} a_i r^i \qquad (24)$$

Here, for finding the constant coefficients $a_i$, first use the problem's boundary conditions for chosen answer. Then for providing more equations for finding the constant coefficients, we substitute the chosen answer into the differential equation as follows:

$$f(\frac{d^2(\sum_{i=0}^{n} a_i r^i)}{dr^2}, \frac{d(\sum_{i=0}^{n} a_i r^i)}{dr}, (\sum_{i=0}^{n} a_i r^i), q_r) = 0$$
(25)

Now, the differential equation changes to a algebraic equation respect to independent variable $r$. so, we have:

$$g(r) = f(\frac{d^2(\sum_{i=0}^{n} a_i r^i)}{dr^2}, \frac{d(\sum_{i=0}^{n} a_i r^i)}{dr}, (\sum_{i=0}^{n} a_i r^i), q_r)$$
(26)

Then, to provide required equations for find constant coefficients $a_i$, we use last equation in the following form:

$$\frac{d^m}{dr^m}(g(0)) = 0$$
(27)

Where m is from 0 to (n-'number of boundary conditions').

Now, unknown constant coefficients can be easily calculated by solving a set of algebraic equations which is consisted of n equations and n unknowns.

Just as noted previously, the chosen answer is as follows:

$$\theta(r) = r^3 a_3 + r^2 a_2 + r^1 a_1 + a_0 \qquad (28)$$
$$\phi(r) = r^3 b_3 + r^2 b_2 + r^1 b_1 + b_0 \qquad (29)$$
$$V(r) = r^3 c_3 + r^2 c_2 + r^1 c_1 + c_0 \qquad (30)$$

By following the steps of AGM, When
$\alpha = 1, N_b = 2, N_t = 2, A = 1, M = 1, G_r = 1, B_r = 1, P = 1, c = -1$

finally 12 unknown constant coefficients calculated and the answers are:

$$\theta(r) = -0.1199019387 r^3 - 0.06933840624 r^2 \\ + 0.04732878977 r + 1.141911555 \qquad (31)$$

$$\phi(r) = -0.1810698649 r^3 + 2.092059187 r^2 \\ - 6.008688506 r + 5.097699184 \qquad (32)$$

$$V(r) = -0.01831262236 r^3 + 0.3810640681 r^2 \\ - 2.015003848 r + 2.652252402 \qquad (33)$$

Table 1. some properties of non-Newtonian fluid and nanoparticles

| Material | Symbol | Density (kg/m3) | Cp (J/kg k) | Thermal conductivity, k (W/m k) |
|---|---|---|---|---|
| Gold | Au | 19300 | 129 | 318 |
| Blood | - | 1050 | 3617 | 0.52 |

Table 2. Comparison of values of V(r) obtained by AGM with numerical solution, GM, CM and LSM when
$\alpha = 1, N_b = 2, N_t = 2, A = 1, M = 1, G_r = 1, B_r = 1, P = 1, c = -1$

| | Numeric | | AGM | | Error | |
|---|---|---|---|---|---|---|
| r | θ(r) | φ(r) | θ(r) | φ(r) | θ(r) | φ(r) |
| 1 | 1 | 1 | 1 | 1 | 0 | 0 |
| 1.1 | 0.950455 | 0.77453 | 0.950481 | 0.77852 | 0.000025 | 0.00399 |
| 1.2 | 0.891648 | 0.58228 | 0.891663 | 0.58694 | 0.000014 | 0.00466 |
| 1.3 | 0.822834 | 0.42014 | 0.822825 | 0.42417 | 0.000009 | 0.00403 |
| 1.4 | 0.743280 | 0.28586 | 0.743248 | 0.28911 | 0.000031 | 0.00324 |
| 1.5 | 0.652256 | 0.17781 | 0.652214 | 0.18068 | 0.000042 | 0.00287 |
| 1.6 | 0.549041 | 0.94814 | 0.549003 | 0.97806 | 0.000038 | 0.02992 |
| 1.7 | 0.432918 | 0.36012 | 0.432895 | 0.39383 | 0.000022 | 0.03371 |
| 1.8 | 0.303173 | 0.83237 | 0.303171 | 0.84332 | 0.000002 | 0.01094 |
| 1.9 | 0.159102 | -0.11100 | 0.159113 | -0.18433 | 0.00001 | 0.07332 |
| 2 | 0 | 0 | 0 | 0 | 0 | 0 |

Table 3. Comparison of values of θ(r) and ϕ(r) obtained by AGM with numerical solution when
$\alpha=1, N_b=2, N_t=2, A=1, M=1, G_r=1, B_r=1, P=1, c=-1$

| r | V(r) Numeric | AGM | Error GM [17] | Error LSM [17] | Error CM [17] | Error AGM |
|---|---|---|---|---|---|---|
| 1   | 1        | 1        | 0       | 0       | 0       | 0       |
| 1.1 | 0.871765 | 0.872461 | 0.00176 | 0.00394 | 0.00226 | 0.00069 |
| 1.2 | 0.750453 | 0.751335 | 0.00599 | 0.01002 | 0.00567 | 0.00088 |
| 1.3 | 0.635675 | 0.636512 | 0.01255 | 0.01806 | 0.01013 | 0.00083 |
| 1.4 | 0.521746 | 0.527882 | 0.02124 | 0.0279  | 0.01554 | 0.00613 |
| 1.5 | 0.424654 | 0.425335 | 0.03182 | 0.03925 | 0.02179 | 0.00068 |
| 1.6 | 0.328059 | 0.328761 | 0.0439  | 0.05174 | 0.02873 | 0.00070 |
| 1.7 | 0.237282 | 0.238051 | 0.05687 | 0.05174 | 0.03617 | 0.00076 |
| 1.8 | 0.152305 | 0.153093 | 0.06962 | 0.07719 | 0.04388 | 0.00078 |
| 1.9 | 0.073172 | 0.073780 | 0.07986 | 0.08674 | 0.05157 | 0.00060 |
| 2   | 0        | 0        | 0       | 0       | 0       | 0       |

In this study viscosity of Nanofluid is determined by using Constant Model, Reynolds' Model and Vogel's Model [3],

$\mu = 1$ (34) $\Rightarrow$ $\mu = e^{-b\theta}$ (35) $\Rightarrow$ $\mu = \mu_0 e^{\frac{a}{b+\theta}\theta_0}$ (36)

Where a and b are constants, θ is temperature and subscript "0" denotes to the ambient condition.

IV. RESULTS AND DISCUSSION

In the present study, the behavior of blood in arteries as a third grade non-Newtonian fluid is investigated analytically. Akbari-Ganji's Method (AGM) is applied to solve the nonlinear differential equations and the results are also compared with numerical approach. Fig. 2 shows the accuracy and validation of AGM in comparison with fourth order Runge-Kutta numerical solution. This figure showed that AGM has an excellent agreement with numerical results and can be introduced as a simple and powerful analytical method for these kinds of problems.

As the Table 2 indicates, the precision of AGM is even more than GM, CM and LSM. Also Table 3 indicate the accuracy of AGM in comparison with numerical solution for temperature and concentration of nanoparticles. Also result with $\alpha=1, N_b=2, N_t=2, A=1, M=1, G_r=1, B_r=1, P=1, c=-1$ was concluded and reported.

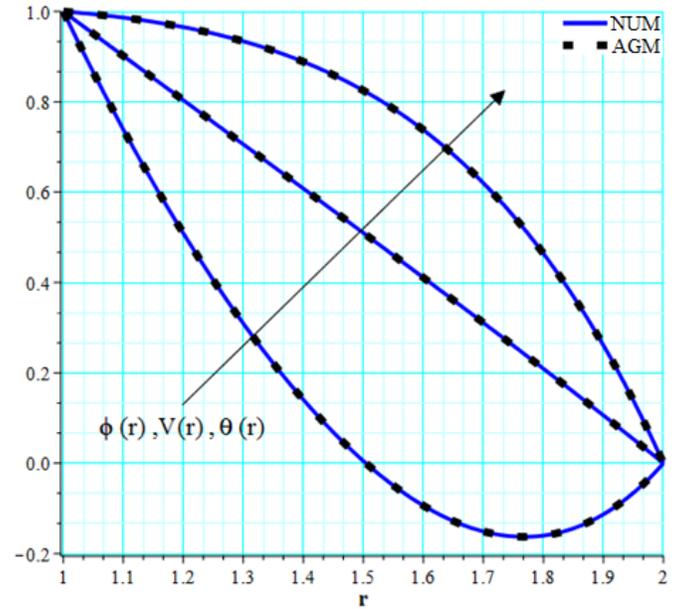

Figure 2. comparison between AGM and numerical method for nanoparticle concentration, velocity and temperature profile

Fig. 3 displays the effect of pressure gradient (c) on the velocity. The pressure gradient developed between the arterial and the venous end of the circulation is the driving force causing blood flow through the vessels which is negative across the aortic valve throughout the cardiac cycle. As the figure shows, an increase of pressure gradient causes an increase of the velocity profile that contributes to an increase of the blood velocity in arteries.

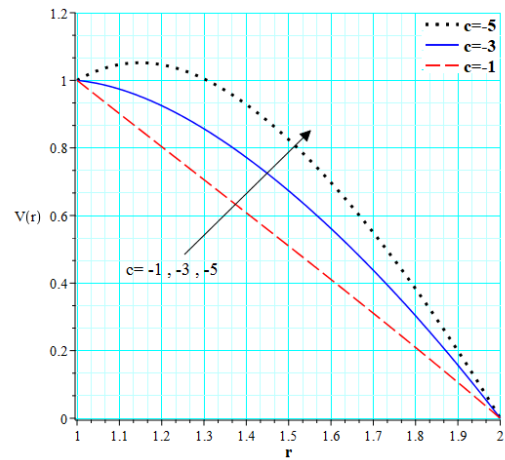

Figure 3. Effect of pressure gradient parameter on velocity profile

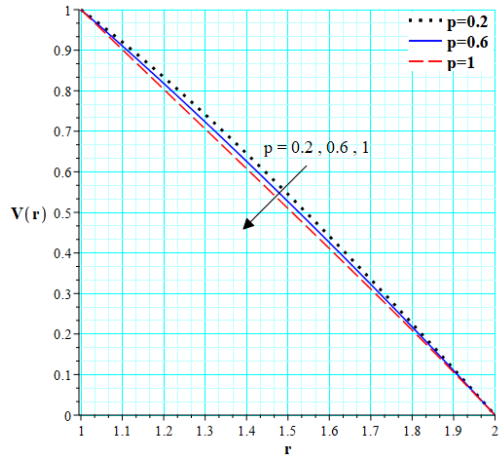

Figure 4. Effect of porosity parameter on velocity profile

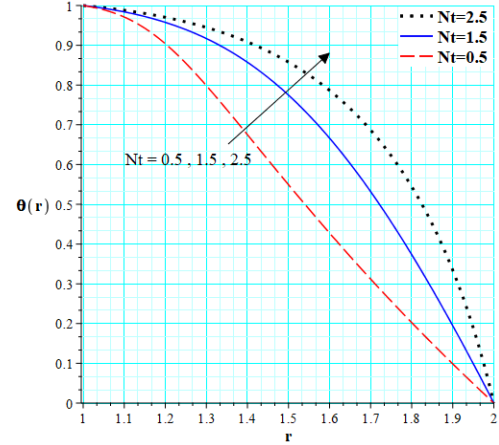

Figure 7. Effect of thermophoresis parameter on temperature profile

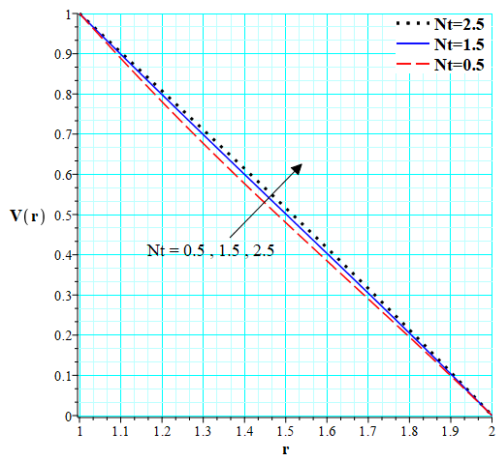

Figure 5. Effect of thermophoresis parameter on velocity profile

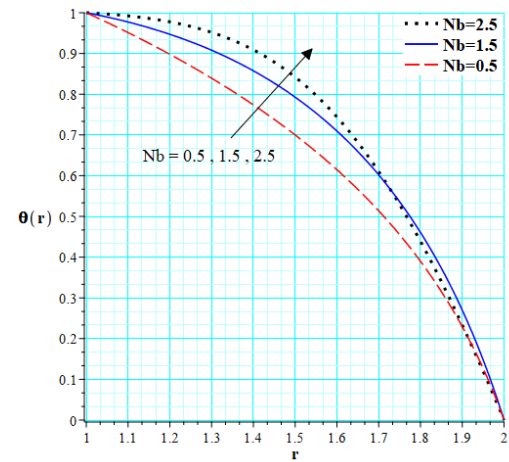

Figure 8. Effect of Brownian motion parameter on temperature profile

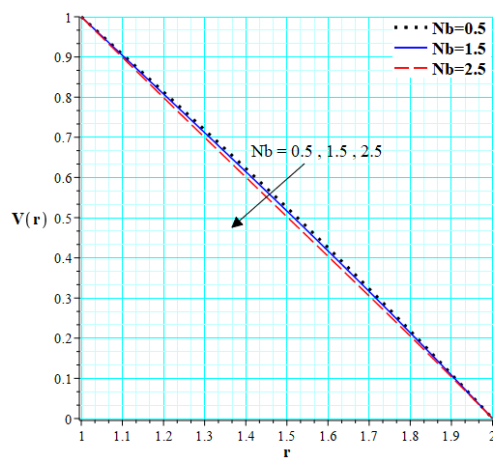

Figure 6. Effect of Brownian motion parameter on velocity profile

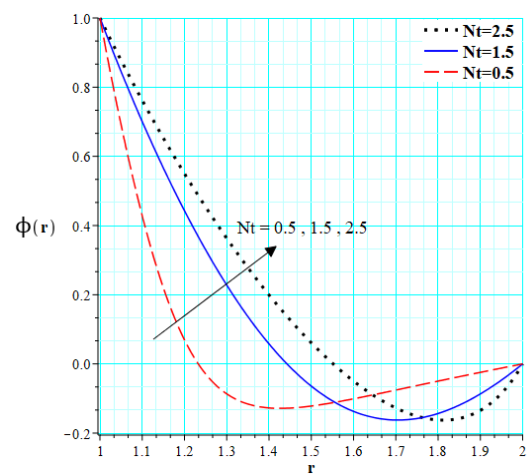

Figure 9. Effect of thermophoresis parameter on concentration of nanoparticles profile

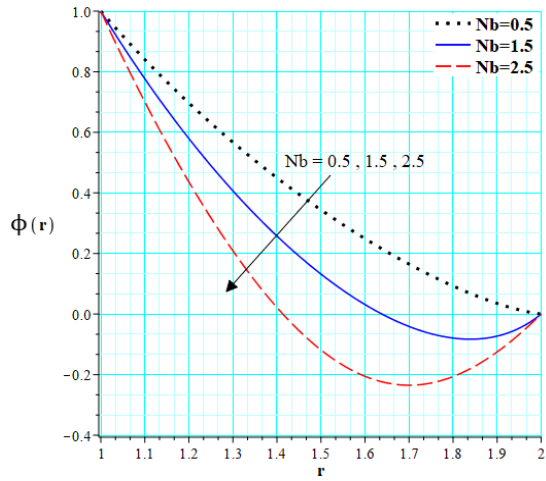

Figure 10. Effect of Brownian motion parameter on concentration of nanoparticles profile

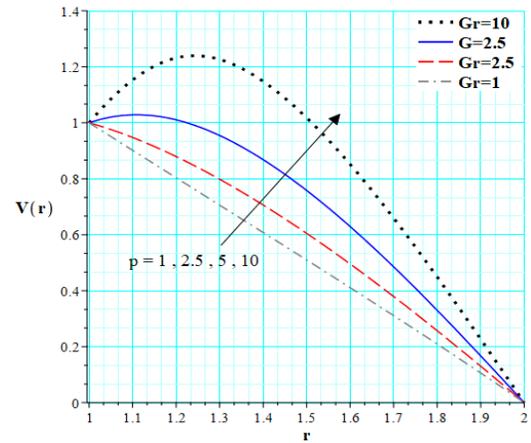

Figure 13. Effect of Grashof number parameter on velocity profile

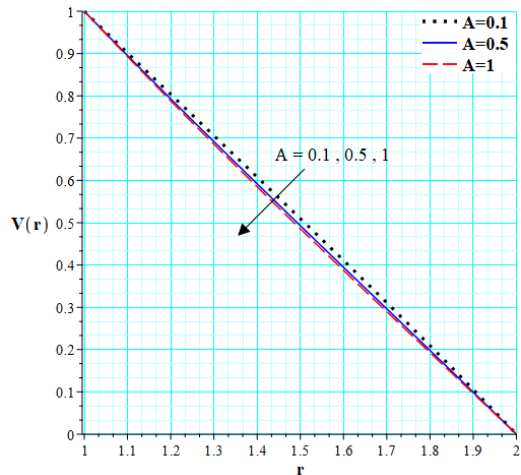

Figure 11. Effect of third grade parameter on velocity profile

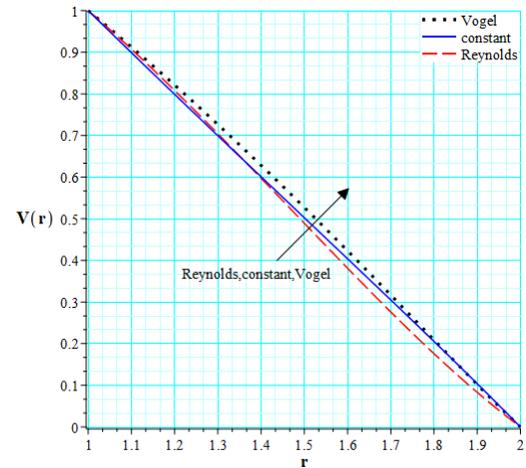

Figure 14. Comparison constant, Reynolds and Vogel's viscosity model for velocity profile

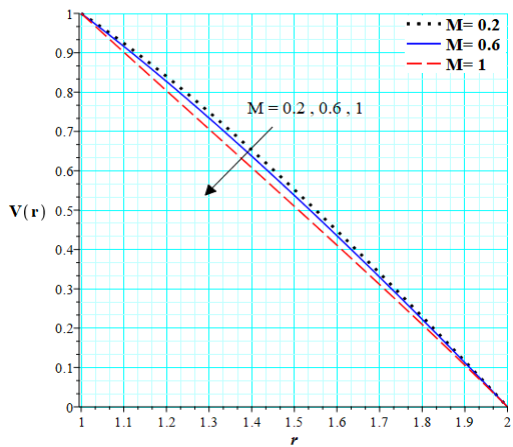

Figure 12. Effect of third grade parameter on velocity profile

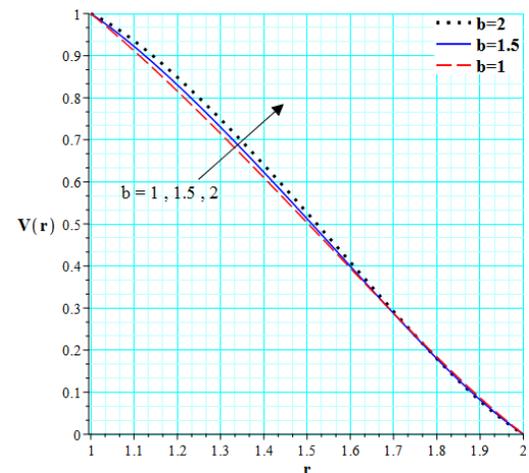

Figure 15. Effect of Reynolds viscosity power index on velocity profile

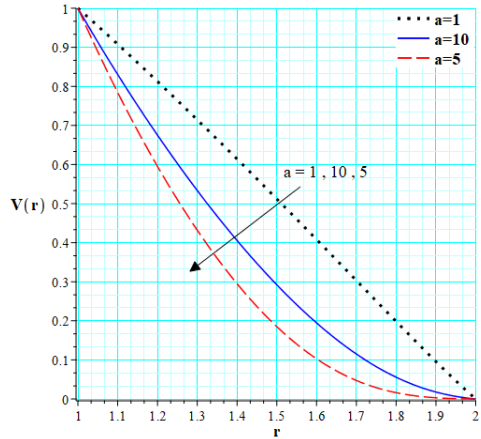

Figure 16. Effect of Vogel's viscosity power index "a" on velocity profile

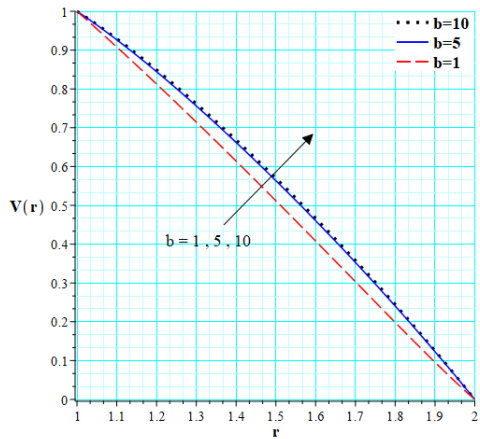

Figure 17. Effect of Vogel's viscosity power index "b" on velocity profile

As known porosity and pore size distributions can be tuned to enhance the transformation of various Nano-medicines. Therefore, understanding the effect of porosity parameter is of great importance. Fig. 4 illustrates the velocity profile distribution across various values of porosity parameter that velocity profile slightly increases with reduction in porosity. Because when porosity increases, fluid can easily move between the pores and as a result velocity profiles increases.

The effect of thermophoresis parameter and Brownian motion parameter on velocity profile is indicated in Fig. 5 and Fig. 6. According to these figures, an increase of $N_t$ and decrease of $N_b$ lead to gradual increase in the velocity profile. Fig. 7 shows the influence of thermophoresis parameter on non-dimensional temperature profile. It is obvious that the temperature profile is an increasing function of $N_t$ number. From the physical point of view an increase in the thermophoresis number can generate higher amount of mass flux due to temperature gradient which consequently rises the nanoparticle concentration. Fig. 8 illustrates the effect of Brownian motion parameter on the temperature profile. Brownian motion plays a vital role in affecting the flow path of platelets or any other cells. This parameter can be negative or positive. Negative value of it indicates that the concentration of nanoparticles on the wall surface is less than that outside the boundary layer. This figures confirms that Brownian motion parameter doesn't have a predictable behavior. At values of $N_b = 1.5$ and $N_b = 2.5$, the temperature profile crosses each other when $0 < \theta < 0.5$ and it overlaps at $0.5 < \theta < 1$. Apart from this observation at lower $N_b$, the temperature and boundary layer thickness show a reasonable trend at higher Brownian motion number. As seen, Brownian motion number and thermophoresis number don't notable effect on temperature profile that we expect it.

The effects of thermophoresis parameter and Brownian motion parameter on nanoparticle concentration of the Nanofluid are displayed in Fig. 9 and Fig. 10. According to Fig. 9, the concentration of nanoparticles has a direct impact on $N_t$ and it enhances by increasing the thermophoresis number. However, the opposite trend is observed in Fig. 10 implying that by increasing the $N_b$, nanoparticle concentration of the Nanoluid decreases. This is probably due to the fact that the impact of thermal conductivity increases by enhancement in Brownian motion which leads to the reduction in the concentration profile. It can be concluded that when the particle size limits to the Nano-sized scale medicine, the Brownian motion and its impact on the surrounding fluid play a significant role in the heat transfer. As seen, Brownian motion number and thermophoresis number don't notable effect on concentration of nanoparticles profile that we expect it.

The effect of the third grade parameter on velocity profile is depicted in Fig. 11. It is evident that an increase of the A number contributes to an increase in the velocity profile. The effect of MHD parameter M on the velocity profile is shown in Fig. 12. As known, blood is considered as magnetic fluid, in which red blood cells are magnetic in nature. Liquid carriers in the blood contain the magnetic suspension of the particle. It is observed that the velocity reduces at higher values of magnetic field intensity. The main reason for this trend is that when the magnetic field is applied, the flow of blood is opposed as a result of the Laurentz force. Therefore, the magnetic field exposure slows down the blood velocity.

Fig. 13 illustrates the effect of Gr on the velocity profile. As seen, increasing Gr speeds up the velocity of the flow field at all points because of the action of convection current on the flow field. Fig. 14 depicts the velocity profile as a function of particle radius for different Nanofluid viscosity models (Constant, Reynolds' and Vogel's models). While Reynolds and Vogel models are applicable for the viscosity of a third-grade Nanofluid, the constant viscosity model can only be applied to Newtonian fluid and is not a reasonable assumption for blood flow. It is elucidated that the highest values of velocity are achieved when the Vogel's model is used for viscosity of Nanofluid.

Finally the effect of Reynolds viscosity power index b on velocity profile (32) is showed in Fig. 15. As seen, by increasing b, velocity in most point of domain increased. Also the effect of constant numbers appeared in Vogel viscosity models (33), *a* and *b*, is demonstrated in Fig. 16 and Fig. 17. It is obvious that by increasing the *a* parameter, velocity at first (to a=5) decreases then increases, but by increasing the b parameter, velocity profile increases slightly.

V. CONCLUSION

The performance of the biological systems can be improved by controlling the nanoparticle behavior. This allows the modulation of immune system interactions, interaction with target cells, and consequently helping in the effective delivery of cargo within cells or tissues. In this paper, blood is modeled as third-grade non-Newtonian flow and nanoparticles in the presence of magnetic field added to it. governing nonlinear differential equations are solved by Akbari-Ganji's Method (AGM) and compared with numerical approach (Runge-Kutta method). The obtained results reveal that AGM achieves suitable accuracy in predicting the solution of such problems in comparison with other same works. it is concluded that the increase in pressure gradient and Grashof number with the decrease in magnetic field intensity accelerate the Nano-fluid's velocity and the highest velocity is achieved by the Vogel's model for viscosity. Moreover, the Brownian motion and thermophoresis play a vital role on the temperature and nanoparticle concentration distributions. either increase in Thermophoresis or decrease in Brownian motion parameters results in enhancement in nanoparticle concentration. Also the effect of constant numbers appeared in Vogel's viscosity model was reported that by increasing *a* parameter, velocity at first (to a=5) decreases then increases, but by increasing the *b* parameter, velocity profile increases slightly.

Nomenclature

| | |
|---|---|
| $CM$ | Collocation Method |
| $GM$ | Galerkin Method |
| $LSM$ | Least Square Method |
| $AGM$ | Akbari-Ganji's Method |
| $WRM$ | Weighted Residuals Methods |
| $c$ | pressure gradient |
| $D_b$ | Brownian diffusion coefficient |
| $D_T$ | thermophoretic diffusion coefficient |
| $g$ | gravitational vector |
| $G_r$ | Thermophoresis diffusion constant |
| $a, b$ | constants in viscosity function |
| $\kappa$ | permeability |
| $M$ | MHD parameter |
| $W(x)$ | weighted function |
| $A$ | third-grade parameter |
| $R(x)$ | residual function |
| $V$ | velocity vector |
| $B_r$ | Brownian diffusion constant |
| $N_b$ | Brownian motion parameter |
| $N_t$ | thermophoresis parameter |
| $P$ | porosity parameter |
| $R_1$ | inner radius |
| $R_2$ | outer radius |
| *Greek symbols* | |
| $\alpha, \beta$ | material moduli |
| $\lambda_r$ | retardation time |
| $\mu$ | viscosity |
| $\phi$ | nanoparticle volume fraction |
| $\theta$ | temperature |
| $\rho_f$ | density of the base fluid |
| $\rho_p$ | density of the nanoparticles |
| $\varphi$ | porosity |
| $\beta_T$ | volumetric expansion coefficient |